\documentclass[12pt,a4paper,oneside]{article}
\usepackage{mathrsfs}
\usepackage[hmargin=2.5cm,vmargin=3cm]{geometry}
 \usepackage{graphicx}
\usepackage{amsmath,amssymb,amsthm,mathtools}
  \usepackage{enumerate}
\usepackage{cite}
\usepackage{mathabx}
\usepackage{url}
\usepackage{hhline}
\newcommand{\bey}{\begin{eqnarray}}
\newcommand{\eey}{\end{eqnarray}}
\usepackage{etoolbox}
\usepackage{accents}
\usepackage{bm}
\newcommand{\uotimes}{\mathbin{\underaccent{\bar}{\otimes}}}
\theoremstyle{definition}
\newcommand{\po}{\prec}

\let\OLDthebibliography\thebibliography
\renewcommand\thebibliography[1]{
  \OLDthebibliography{#1}
  \setlength{\parskip}{0pt}
  \setlength{\itemsep}{2pt plus 0.3ex}
}
 \usepackage{enumitem}
\setlist[itemize]{leftmargin=1.2em}
\setlist[enumerate]{leftmargin=1.2em}

\newcommand{\R}{\mathbb{R}}

\begin{document}
\title{Superposition of dynamics, indefinite causal order, and quantum histories }
\author { Charis Anastopoulos\footnote{anastop@upatras.gr} \;   and   \; Ntina Savvidou\footnote{ksavvidou@upatras.gr}\\
 {\small Laboratory of Universe Sciences, Department of Physics, University of Patras, 26500 Greece} }
\maketitle

 \begin{abstract}
The process-matrix formalism describes quantum processes without assuming a fixed global causal order, but the physical meaning of indefinite causal order remains open. We address this question within histories theory, where temporal ordering and dynamical evolution are distinct structures. We show that the decoherence functional admits coherent superpositions of dynamics. In measurement settings, under suitable factorization conditions, these generate process matrices. In the bipartite setting considered here, process-matrix indefinite causal order is thereby identified as an operationally restricted realization of the more general phenomenon of superposition of dynamics.  Histories theory also admits a distinct kinematical notion of indefinite order, in which the ordering of physical events is itself a history observable. The resulting distinction between event order and intervention order clarifies the relation between indefinite causal order, quantum dynamics, and spacetime structure.
 
 \end{abstract} 

 \section{Introduction}

The process-matrix formalism \cite{OreshkovCostaBrukner2012, ChiribellaDArianoPerinottiValiron2013, Araujo15, OrGi16, CoSh16} provides an operational framework for assigning probabilities to quantum operations without assuming that these operations are embedded in a fixed global causal order. Processes of this kind are commonly described as exhibiting indefinite causal order (ICO), or a quantum superposition of causal orders. Despite substantial theoretical and experimental progress, however, the physical content of this notion remains open. Recent assessments have highlighted  questions concerning the status of events and the relation between operationally defined causal relations and the causal structure of spacetime \cite{Costa26}.

In this paper, we approach these questions from the perspective of histories theory \cite{Gri84,Gri,Omn1,Omn2,GeHa1,hartlelo,Sorkin1,Ish94,Sav10}. Histories theory is a family of formulations of quantum theory built around temporally extended alternatives, or histories, rather than single-time states or observables. The decoherent-histories approach is its best-known representative. In these formulations, probabilities are encoded in the decoherence functional, a bilinear function of pairs of histories. Its off-diagonal terms quantify interference between alternative histories.

Particularly relevant here is the dual-time histories theory developed by one of us (K.S.) \cite{Sav99,Sav10, Sav1, Sav2}. It reflects the existence of two distinct types of time transformation: a kinematical one associated with the temporal ordering of history propositions, and a dynamical one associated with the assignment of probabilities. This distinction makes the histories framework especially well suited to separating different notions of temporal and causal order, and provides the structural basis for the analysis developed below.

Our motivation is therefore to disentangle the different conceptual strands entering the notion of ICO and to embed it in a more general, non-operational framework that can connect naturally with fundamental physics. To this end, we reformulate the process-matrix formalism in the language of histories theory and establish a direct relation between the two frameworks at the level of the decoherence functional. Process matrices can be identified with effective strongly positive decoherence functionals of the measured degrees of freedom, while ICO corresponds to particular non-canonical structures within this broader histories framework.

This analysis leads naturally to a more general concept, which we call a \emph{superposition of dynamics}. It refers to a structure of the decoherence functional in which distinct dynamical contributions---associated, for example, with different Hamiltonians---interfere coherently. Such a structure goes beyond a statistical mixture of dynamics, which is already allowed by classical probability theory \cite{Str09}. It is intrinsically quantum and, in general, cannot be represented within a description based solely on single-time quantum states and their evolution.

We show that, in an operational setting, a process matrix can be identified with the effective decoherence functional of the measured degrees of freedom. Conversely, when suitable factorization conditions are satisfied, decoherence functionals containing superpositions of dynamics generate process matrices. Within this histories representation, ICO corresponds to superpositions of dynamics that encode different sequences of interaction between the measured system and the measuring apparatuses. The explicit results presented here concern bipartite systems, but the structure of the construction gives strong reason to expect that the correspondence extends to the general multipartite case.

Histories theory also admits a conceptually different realization of   ICO. One may define histories whose alternatives correspond directly to different causal orderings of events \cite{AnPl23}, where by an ``event'' we mean a definite macroscopic measurement record. The causal order of such events is then itself a history observable and, like any other quantum observable, may exhibit interference between its alternatives.

\begin{figure}[h]
 \includegraphics[width=0.8\textwidth]{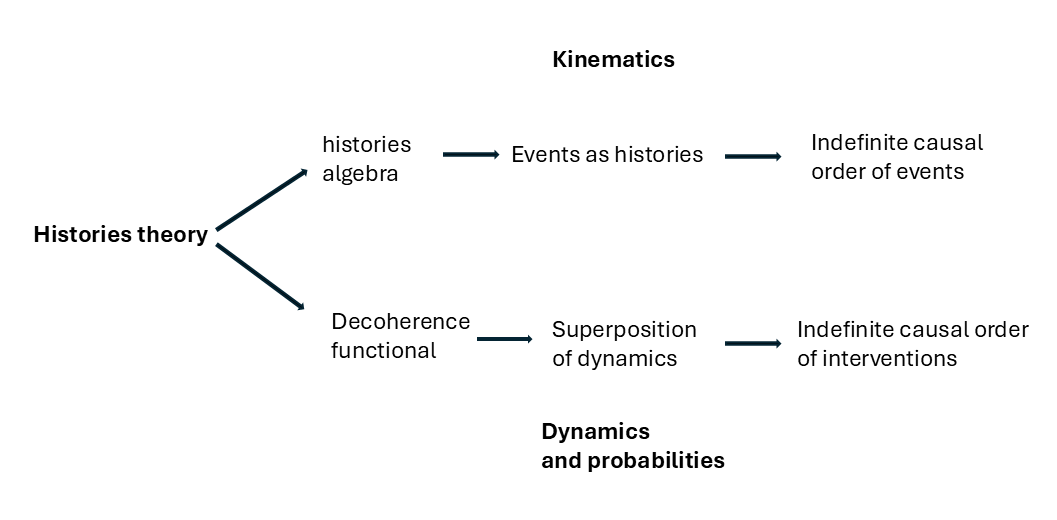}
  \caption{ Two notions of indefinite causal order in histories theory.
}
      \label{flow}
\end{figure}

We are thus led to distinguish two notions of ICO, reflecting the dual-time structure of histories theory:
\begin{enumerate}
\item A \emph{kinematical} notion, in which the causal ordering of events is itself a history variable.
\item A \emph{dynamical} notion, arising as a special case of superposition of dynamics, in which different dynamical contributions correspond to different sequences of interventions.
\end{enumerate}
The underlying structural basis   is shown in Fig. \ref{flow}.

The distinction above also reflects two different uses of ``causal'' in physics. In one sense, causality concerns temporal ordering constrained by the light-cone structure of spacetime; in another, it concerns causal influence (causation), encoded in the sequence of interactions among physical systems. Indefinite causal order in the process-matrix formalism belongs naturally to the second category. It therefore does not, by itself, imply a superposition of spacetime causal structure and can be compatible with a fixed background causal structure \cite{PaVo20, ViRe}.  
  
 \section{Basic notions of histories theory}
\subsection{Definitions}

Classical physics admits both an evolutionary description and a histories description. In the former, the basic object is a microstate that evolves in time, as in Hamiltonian mechanics. In the latter, the basic objects are the possible histories, or trajectories, together with a principle that selects the physically realised ones, namely, the action principle.

Quantum theory is usually formulated in Hamiltonian, and hence evolutionary, terms. A histories formulation is more subtle. Its starting point is the notion of a \emph{history}: a sequence of properties of a physical system at successive instants of time. An $N$-time history $\alpha$ is represented by a sequence
\[
\alpha_{t_1},\alpha_{t_2},\ldots,\alpha_{t_n}
\]
of projection operators. Typically, each projector $\alpha_{t_i}$ represents a particular outcome in a measurement of an observable $\hat A^{(i)}$ at time $t_i$, for $i=1,2,\ldots,n$.

To assign a quantum probability to a history, we introduce the associated class operator
\bey
C_{\alpha} = \alpha_{t_1}(t_1)\alpha_{t_2}(t_2)\cdots\alpha_{t_n}(t_n), \label{clop}
\eey
where
$$
 \alpha_t(t)=e^{i Ht}\alpha_t e^{-i  Ht}
$$
is the Heisenberg-picture evolution of $\alpha_t$. The probability of the history $\alpha$ is then
\begin{equation}
p(\alpha)=\langle\psi|  C_\alpha  C_\alpha^\dagger|\psi\rangle. \label{histprob2}
\end{equation}
Equation (\ref{histprob2}) was proposed by Wigner and collaborators as expressing the essential empirical content of quantum theory \cite{Houtapel}, since it provides the probabilities associated with any conceivable sequence of quantum measurements.

The decoherent-histories approach, developed by Griffiths, Omn\'es, Gell-Mann, and Hartle \cite{Gri84, Gri, Omn1, Omn2, GeHa1, GeHa93, hartlelo}, reinterprets a history as a sequence of properties possessed by a physical system rather than as a sequence of measurement outcomes. This reinterpretation gives histories a powerful logical structure: one may formulate propositions about histories and combine them by logical operations.

The difficulty is that Eq. (\ref{histprob2}) does not, in general, satisfy the Kolmogorov additivity condition
\bey
p(\alpha\vee\beta)=p(\alpha)+p(\beta) \label{eqkol}
\eey
for arbitrary disjoint histories $\alpha$ and $\beta$.

A partial resolution is obtained by restricting attention to suitable sets of histories. To formulate this restriction, one introduces the \emph{decoherence functional} $d$, a complex-valued function of pairs of histories, defined by \index{decoherence functional}
\begin{equation}
d(\alpha,\beta)=\operatorname{Tr}\!\left( C_\alpha^\dagger \rho_0  C_\beta\right). \label{decfun}
\end{equation}
Its diagonal elements coincide with the probabilities in Eq. (\ref{histprob2}),
\[
d(\alpha,\alpha)=p(\alpha).
\]

Let $\Omega$ be an exclusive and exhaustive set of histories: its elements are mutually disjoint and their disjunction is the unit history. If every pair of distinct histories in $\Omega$ satisfies the decoherence condition
\begin{equation}
d(\alpha,\beta)=0,\qquad \alpha\neq\beta, \label{decc}
\end{equation}
then Eq. (\ref{eqkol}) holds, and a probability measure can be defined on $\Omega$. Such a set is called a \emph{consistent set} or a \emph{framework}. Strictly speaking, the weaker condition
\[
\operatorname{Re}d(\alpha,\beta)=0,\qquad \alpha\neq\beta,
\]
is sufficient for consistency. The stronger condition (\ref{decc}), however, is generally better suited to discussions of the classical limit.

\subsection{HPO formalism}

Gell-Mann and Hartle proposed an axiomatic formulation in which a history is treated as an irreducible entity in its own right, rather than necessarily as a time-ordered string of single-time propositions. Isham subsequently generalised and strengthened this framework through the History Projection Operator (HPO) formalism \cite{Ish94, IL94, ILS}, in which histories are represented by projection operators on an appropriate Hilbert space.

We now summarise the basic structure of the Gell-Mann--Hartle--Isham (GHI) axiomatic framework.

\subsubsection{The space of histories}

The set $\mathcal U$ contains all possible histories, generalizing the sequence of projectors of the standard formulation. The set of histories $\mathcal U$ is equipped with a \emph{partial ordering} $\leq$ for coarse-graining.  The relation $\alpha\leq\beta$ means that $\beta$ is a coarser description of the same physical situation than $\alpha$.

The set $\mathcal U$ contains the   \emph{unit history} $I$, corresponding to the unrestricted proposition that ``something happens''; and the 
\emph{null history} $0$, corresponding to logical impossibility. All histories satisfy $0\leq\alpha\leq I$.
 
A history $\alpha$ is \emph{fine-grained} if the only histories $\beta$ satisfying $\beta\leq\alpha$ are $0$ and $\alpha$ itself.

%\item \emph{Logical AND}.
%There is an operation $\wedge$ representing logical conjunction. The history $\alpha\wedge\beta$ represents the proposition that both $\alpha$ and $\beta$ hold.

%\item \emph{Temporal composition}.
%Histories may also be composed in time. If it is meaningful to say that $\beta$ occurs after $\alpha$, one may form the composite history $\alpha\circ\beta$. In this case, $\beta$ follows $\alpha$, or equivalently $\alpha$ precedes $\beta$.

%For homogeneous histories, temporal composition corresponds to concatenating two time-ordered sequences. Whenever it is defined, it agrees with logical conjunction. In temporal logic, $\circ$ represents the connective AND THEN.

\subsubsection{The space of temporal supports}

The logical structure of histories is  supplemented by a temporal structure. This is encoded in a second space, denoted by $Sup$, whose elements are the \emph{temporal supports} of histories. For a homogeneous history $\alpha:= (\alpha_{t_1},\alpha_{t_2},\ldots,\alpha_{t_n})$, the temporal support is the set $\sigma(\alpha) = \{t_1,t_2,\ldots,t_n\} \subset \R$. 

More generally, a temporal support may be an open subset of the real line $\R$, or may even involve exotic temporal structures. The existence of  
  a map $\sigma:\mathcal U\longrightarrow Sup$, which assigns a temporal support to each history, is assumed axiomatically.
 
Crucially, 
$Sup$ is equipped with a  partial order $\po$ for temporal precedence.  This induces a temporal partial order $\po$ on the space of histories ${\cal U}$: 
$$
\alpha \po \beta, \; \mbox{iff} \;  \sigma(\alpha) \po \sigma(\beta).
$$

%Some histories can be constructed by composing simpler ones,
%\[
%\alpha=\alpha^1\circ\alpha^2\circ\cdots\circ\alpha^N. \label{dechis}
%\]
%A history that admits no such decomposition is called \emph{nuclear}. Similarly, a support $s$ is nuclear if it cannot be decomposed into a sequence of two supports.
%In physical terms, nuclear histories generalise single-time propositions, while nuclear supports generalise individual moments of time.

%Finally, a decomposition such as (\ref{dechis}) is said to be \emph{irreducible} if every component $\alpha^i$ is nuclear. For homogeneous histories, this corresponds to expressing a history as a sequence of single-time propositions ordered in time.

\subsubsection{The space of history propositions}

The space $\mathcal U$ is not sufficient to represent every meaningful statement about a physical system, since one must also consider logical combinations of histories. For this reason, $\mathcal U$ is embedded in a larger space $\mathcal{UP}$ of \emph{history propositions}.

The space $\mathcal{UP}$ has the structure of a logical lattice, with operations
  $\wedge$ (AND),
  $\vee$ (OR), and 
  $\neg$ (NOT).
Every element of $\mathcal{UP}$ can be constructed from elements of $\mathcal U$ by a   sequence of these logical operations. Thus, $\mathcal{UP}$ represents the full space of propositions about histories.

Two history propositions $\alpha$ and $\beta$ are \emph{disjoint}, denoted by $\alpha\perp\beta$, if $\alpha\leq\neg\beta$,
so that they cannot both occur. A set $\{\alpha_i\}$ of propositions is \emph{exclusive} if its elements are pairwise disjoint, and \emph{exhaustive} if their disjunction is the unit proposition, $\vee_i \alpha_i = I$.
Such sets represent mutually exclusive and collectively exhaustive alternatives to which probabilities may be assigned.

A particularly important realisation of this structure represents history propositions by projection operators on a Hilbert space. For  fixed times $\{t_1,\ldots,t_n\}$, a history $\alpha=(\alpha_{t_1},\alpha_{t_2},\ldots,\alpha_{t_n})$
is represented by the tensor-product operator
\begin{equation}
\tilde\alpha:=\alpha_{t_1}\otimes\alpha_{t_2}\otimes\cdots\otimes\alpha_{t_n},
\label{Def:th_clean}
\end{equation}
acting on the tensor-product Hilbert space
\bey
{\cal V}=\mathcal H_{t_1}\otimes\mathcal H_{t_2}\otimes\cdots\otimes\mathcal H_{t_n}. \label{calv}
\eey
Hence ${\cal UP}$ is identified with the lattice $\mathcal P({\cal V})$ of projector operators on ${\cal V}$.
This is the \emph{history projection operator} (HPO) representation. Logical operations on history propositions are thereby realised as operations on projection operators, as in the standard mathematical framework of quantum theory. 
However, histories theory works fine in reproducing the predictions of quantum theory, even when ${\cal UP}$ is a Boolean algebra, defined on a classical path space \cite{An03}.

The extension from $\mathcal U$ to $\mathcal{UP}$ is essential for a probabilistic theory of histories. Whereas $\mathcal U$ encodes individual possible histories, $\mathcal{UP}$ also describes alternatives, coarse-grainings, and general logical combinations of histories.

\subsubsection{The space of decoherence functionals}

The final ingredient of the histories framework is the \emph{decoherence functional}. It assigns a complex number $d(\alpha,\beta)$ to each pair of history propositions and quantifies the interference between them. The set of all decoherence functionals is denoted by $\mathcal D$.

A decoherence functional contains both the dynamical information of the theory and the probabilistic structure associated with its histories. It is required to satisfy the following conditions.

\begin{itemize}

\item \emph{Hermiticity:} $d(\alpha,\beta)=d(\beta,\alpha)^*$.

\item \emph{Positivity:} $d(\alpha,\alpha)\geq 0$.

\item \emph{Null triviality:} $d(0,\alpha)=0$.

\item \emph{Additivity:} If $\alpha\perp\beta$, then $d(\alpha\vee\beta,\gamma)=d(\alpha,\gamma)+d(\beta,\gamma)$.

\item \emph{Normalisation:} $d(I,I)=1$.

\end{itemize}

The set ${\cal D}$ of decoherence functionals is convex. Thus, if $d_a$ is a finite collection of decoherence functionals and $\lambda_a\in[0,1]$ satisfy $\sum_a\lambda_a=1$, then
\[
d=\sum_a\lambda_a d_a
\]
is also a decoherence functional.

Isham, Linden, and Schreckenberg \cite{ILS} established a generalisation of Gleason's theorem according to which a decoherence functional can be written as
\bey
d(\alpha,\beta)=\operatorname{Tr}(X\,\tilde\alpha\otimes\tilde\beta),
\eey
where $X$ is an operator on ${\cal V}\otimes{\cal V}$ satisfying
\begin{enumerate} 
\item $\operatorname{Tr}_{{\cal V}\otimes{\cal V}}X=1$,
\item $X^\dagger=MXM$, where  $M(\psi\otimes\phi)=\phi\otimes\psi$ is the exchange operator on ${\cal V}\otimes{\cal V}$, and
\item  $\operatorname{Tr}_{{\cal V}\otimes{\cal V}}\!\left[(\tilde\alpha\otimes\tilde\alpha)X\right]\geq0,
\qquad
\forall\tilde\alpha\in\mathcal P({\cal V})$.
\end{enumerate}
We will refer to   $X$ as the ILS operator of the   decoherence functional.

\medskip

\noindent {\em Strong positivity.} Strong positivity strengthens the ordinary positivity axiom for decoherence functionals. Let $d(\alpha,\beta)$ be a decoherence functional and let $\Omega$ be an exclusive and exhaustive set of histories. Strong positivity requires that, for every map $c:\Omega\rightarrow\mathbb C$,
\[
\sum_{\alpha,\beta\in\Omega}c_\alpha c_\beta^*d(\alpha,\beta)\geq0.
\]
Strong positivity allows the decoherence functional to induce an inner product on a history Hilbert space and is preserved under the tensor-product composition of independent systems \cite{DJS10}. For these reasons, it is often regarded as the physically appropriate positivity condition for the decoherence functional \cite{DW22}.

\subsubsection{Histories in continuous time}
To construct the history Hilbert space for continuous time, Isham
et al \cite{IL95, ILSS98} introduced the history group,    
 a generalisation of
the canonical group of standard quantum theory. 
For example, for a particle moving on a line, the single-time canonical
commutation relations,  $[ \hat{x}, \hat{p}  ] = i \hbar$, 
become the history group that is described by the following history commutation relation, defined at unequal moments of time
\begin{eqnarray}
[ \hat{x_t}, \hat{p_{t'}}  ] = i \hbar \delta(t - t').   \label{xtpt'}
\end{eqnarray}

The notion of a ``{\em continuous\/} tensor product"---and hence {\em continuous\/} temporal logic---arises via a representation of the history algebra on a Hilbert space ${\cal V}$. 
Physical quantities are naturally
time-averaged in this scheme; for example, we define the smeared position operator $x_f := \int_{-\infty}^{\infty} x_t f(t)$ on ${\cal V}$ for any square-integrable function $f$. Then, history propositions about position correspond to spectral projectors of the operators $x_f$.

  \subsection{Dual-time histories}
The introduction of the history group in the HPO approach made it possible to define continuous-time histories as physical observables represented by logical propositions. Dynamics was initially absent from this construction, because there was no natural way to relate observables at different time instants, such as $x_t$ and $x_{t'}$.
This problem was resolved through the observation that time enters histories theory in two conceptually distinct ways \cite{Sav99}. First, it appears as the parameter of temporal ordering, distinguishing past, present, and future and thereby encoding the causal ordering of history propositions. Second, it appears as the evolution parameter entering the dynamical laws of a physical system. 

These two roles give rise to distinct types of time transformation:  one associated with temporal ordering, the other with dynamical evolution. For a given physical system, the two transformations are combined through the {\em action operator} $S $, the quantum counterpart of the classical action functional. An analogous distinction exists classically, where the two temporal transformations arise as distinct symplectic transformations on the space of histories \cite{Sav1}.

To illustrate this structure, consider a particle on a line with Hamiltonian function $h(x,p)$. The classical phase-space action is
\begin{equation}
S:= \int^{+\infty}_{-\infty} \! dt \,
\big[p_t\dot{x}_t-h(x_t,p_t)\big].
\label{Def:op_S}
\end{equation}
Its histories analogue is the action operator $\hat S=\hat V-\hat H$ on the continuous-time histories Hilbert space ${\cal V}$. The Liouville operator $\hat V$ represents the kinematical term
$$
V:=\int^\infty_{-\infty}dt\,p_t\dot{x}_t,
$$
while the histories Hamiltonian $\hat H$ represents the time-averaged energy
$$
H=\int_{-\infty}^{\infty}dt\,h(x_t,p_t), 
$$
and generates dynamical evolution. For a smeared history observable $\hat{x}_f$, define its Heisenberg-picture version
\begin{eqnarray}
 x _f(s):=
e^{\frac{i}{\hbar}s H }\,
 x_f\,
e^{-\frac{i}{\hbar}s H} .
\end{eqnarray}
The parameter $s$ is shifted according to
\begin{eqnarray}
e^{\frac{i}{\hbar}\tau H}\,
 x_f(s)\,
e^{-\frac{i}{\hbar}\tau H } = 
 x_f(s+\tau),
\end{eqnarray}
so $s$ is the parameter entering the dynamical laws of the system.

By contrast, the Liouville operator $V$ generates translations of the temporal label entering the history algebra,
\begin{eqnarray}
e^{\frac{i}{\hbar}\tau V }\,
 x_f(s)\,
e^{-\frac{i}{\hbar}\tau V}
=
 x_{f_{\tau}}(s),
\qquad
f_{\tau}(t)=f(t+\tau).
\label{Liouville}
\end{eqnarray}
Hence, $t$ is the parameter of temporal logic: it orders the propositions entering a history without specifying the system's physical evolution.

This distinction between temporal ordering and dynamical evolution is central. The parameter $t$ encodes the ordering structure inherited from physical time, specifying the sequence in which events occur, much as the time-ordering symbol does in quantum field theory. The parameter $s$, by contrast, belongs to the dynamical description and represents physical evolution relative to an internal ``clock". The two therefore represent different aspects of time rather than two copies of the same quantity. The action operator $S $ intertwines these structures and enters the probability assignment and hence the physical predictions of the theory.

The same dual temporal structure persists in relativistic theories. On Minkowski spacetime, the logical ordering of events is expressed through spacetime coordinates $X$, and the two transformations are represented by distinct actions of the Poincar\'e group \cite{Sav02}. An analogous structure arises in classical general relativity, where the histories formulation   accommodates both the spacetime diffeomorphism group and the Dirac algebra of constraints as distinct mathematical objects \cite{Sav2}.

\section{Superposition of dynamics}

In this section, we show that the histories axioms admit decoherence functionals that differ substantially from the standard form (\ref{decfun}). In particular, some of them admit an interpretation in terms of a \emph{superposition of dynamics}, in which different unitary evolution laws contribute coherently to a single decoherence functional.

\subsection{The structure of the decoherence functional}

First, we express the standard decoherence functional in terms of operators on the history Hilbert space ${\cal V}$ of Eq. (\ref{calv}). Define the unitary operators ${\cal S}$ and ${\cal U}$ on ${\cal V}$ by
\bey
{\cal S}|v_1\rangle\otimes|v_2\rangle\otimes\cdots\otimes|v_n\rangle
&=&|v_2\rangle\otimes\cdots\otimes|v_n\rangle\otimes|v_1\rangle,\\
{\cal U}&=&e^{-i Ht_1}\otimes e^{-i Ht_2}\otimes\cdots\otimes e^{-i Ht_n}.
\eey
We refer to ${\cal S}$ as the translation operator and to ${\cal U}$ as the dynamics operator.

For a pure initial state $|\psi_0\rangle$, the decoherence functional (\ref{decfun}) can be written as
\bey
d(\alpha,\beta)=\sum_i c_i^*(\alpha)c_i(\beta), \label{deee2}
\eey
where
\[
c_i(\alpha)=\langle\psi_0|\hat C_\alpha|i\rangle
\]
for any orthonormal basis $\{|i\rangle\}$ of ${\cal H}$.

The amplitudes $c_i(\alpha)$ may be expressed in terms of linear functionals $g_i:{\cal B}({\cal V})\rightarrow\mathbb C$ defined by
\bey
g_i(Y)=\operatorname{Tr}_{\cal V}({\cal A}_iY{\cal S}),
\eey
where
\[
{\cal A}_i=|i\rangle\langle\psi_0|\otimes I\otimes\cdots\otimes I
\]
is the \emph{boundary operator}. It incorporates the initial condition and, when appropriate, may also encode final conditions. With this notation,
\bey
c_i(\alpha)=g_i\!\left({\cal U}^\dagger\tilde\alpha{\cal U}\right),
\eey
and hence
\bey
d(\alpha,\beta)=\sum_i g_i^*\!\left({\cal U}^\dagger\tilde\alpha{\cal U}\right)
g_i\!\left({\cal U}^\dagger\tilde\beta{\cal U}\right).
\eey

The boundary operator can be modified in several ways. For example, an operator of the form
\[
{\cal A}_i=I\otimes\cdots\otimes|i\rangle\langle\psi_0|\otimes\cdots\otimes I
\]
represents conditioning at an intermediate time rather than at the initial time. Post-selected decoherence functionals provide another important case. They factorise as
\[
d(\alpha,\beta)=c^*(\alpha)c(\beta),
\]
with
\[
c(\alpha)=\frac{\langle\psi_f| C_\alpha|\psi_0\rangle}{\langle\psi_f|\psi_0\rangle}
=\operatorname{Tr}_{\cal V}\!\left[{\cal A}{\cal U}^\dagger\tilde\alpha{\cal U}{\cal S}\right],
\]
and
\[
{\cal A}=\frac{1}{\langle\psi_0|\psi_f\rangle}|\psi_0\rangle\langle\psi_f|\otimes I\otimes\cdots\otimes I.
\]
Conditions imposed at several intermediate times can be incorporated in the same way through a suitable choice of boundary operator.

For continuous time, the translation operator ${\cal S}$ is associated to the Liouville operator $V$ of Eq. (\ref{Liouville}),  and depends only on the ordering structure of the underlying spacetime; it is therefore kinematical. By contrast, transformations generated by the operator ${\cal U}$ depend explicitly on the system dynamics.  

\subsection{Non-canonical forms of the decoherence functional}

We define the notion of a superposition of dynamics in terms of the decoherence functional as follows. Fix the functionals $g_i$ and consider amplitudes $c_i(\alpha)$ involving linear combination of terms with different dynamics operators ${\cal U}_a$, labelled by the index $a$,
\bey
c_i(\alpha)=\sum_a b_a g_i\!\left( {\cal U}_a^\dagger\tilde\alpha {\cal U}_a\right), \label{cia}
\eey
where $b_a\in\mathbb C$.

The decoherence functional defined through Eq. (\ref{deee2}) with the amplitudes (\ref{cia}) satisfies the GHI axioms whenever
\[
\left|\sum_a b_a\right| = 1.
\]
The essential point is that the alternatives labelled by $a$ contribute at the amplitude level and therefore interfere.

This construction must be distinguished from a convex combination of dynamics, for which
\bey
d(\alpha,\beta)=\sum_a\lambda_a\sum_i
g_i^*\!\left({\cal U}_a^\dagger\tilde\alpha{\cal U}_a\right)
g_i\!\left({\cal U}_a^\dagger\tilde\beta{\cal U}_a\right),
\eey
where $\lambda_a\in[0,1]$ and $\sum_a\lambda_a=1$. Here the label $a$ identifies mutually exclusive classical alternatives, and no interference occurs between distinct dynamics.

Combining coherent superpositions with convex combinations yields the more general form
\bey
d(\alpha,\beta)=\sum_i\sum_{a,b} \; f_{ab} \;
 g_i^*\!\left( {\cal U}_a^\dagger\tilde\alpha {\cal U}_a\right)
 g_i\!\left( {\cal U}_b^\dagger\tilde\beta {\cal U}_b\right), \label{supdyn}
\eey
where $f_{ab}$ is a positive semidefinite Hermitian matrix satisfying
\[
\sum_{a,b}f_{ab}=1.
\]
Its diagonal entries describe the statistical weights of the different dynamics, whereas its off-diagonal entries encode coherence between them. Note that all decoherence functionals of the form (\ref{supdyn}) satisfy the strong positivity condition.

\medskip

\noindent\textbf{Example.} The most general $2\times2$ matrix satisfying the preceding conditions can be parameterised as
\bey
f=\frac{1}{2(1+n_1)}\left(I+{\bf n}\cdot \bm{\sigma}\right),
\eey
where ${\bf n}=(n_1,n_2,n_3)$ and $|{\bf n}|\leq1$. The most general decoherence functional involving two dynamics, ${\cal U}_1$ and ${\cal U}_2$, is therefore
\bey
d(\alpha,\beta)=\frac{1+n_3}{2(1+n_1)}d_1(\alpha,\beta)
+\frac{1-n_3}{2(1+n_1)}d_2(\alpha,\beta)
+  \frac{n_1-in_2}{2(1+n_1)}F(\alpha,\beta) + \frac{n_1+in_2}{2(1+n_1)}F^*(\beta,\alpha) , \label{superpd}
\eey
where
\bey
F(\alpha,\beta)=\sum_i g_i^*\!\left( {\cal U}_1^\dagger\tilde\alpha {\cal U}_1\right)
g_i\!\left( {\cal U}_2^\dagger\tilde\beta {\cal U}_2\right)
=\langle\psi_0|C_\beta^{(2)}  C_\alpha^{(1) \dagger}  |\psi_0\rangle,
\eey
and $C_\alpha^{(r)}$ is the class operator (\ref{clop}) defined with respect to the dynamics ${\cal U}_r$.

The last two terms define the interference contribution. They vanish  when $f_{ab}$ is diagonal, in which case the construction reduces to a convex combination of the two dynamics. More generally, a decoherence functional containing a genuine superposition of dynamics need not reproduce the single-time structure of standard quantum theory. In particular, even for single-time histories, $d(\alpha,\beta)$ need not be diagonal. Such a functional may therefore yield predictions that cannot be reproduced by standard quantum theory with a single fixed dynamics or by a classical mixture of dynamics.

\medskip

The same construction can be extended beyond the dynamics operator. One may also consider superpositions of the translation operator ${\cal S}$ or of the boundary operator ${\cal A}_i$. Changing ${\cal S}$ changes the order in which the single-time projectors enter the decoherence functional. For a closed system, this ordering is part of the definition of the history space, and ${\cal S}$ is therefore expected to be essentially unique. At the level of an open subsystem, however, different effective translation operators may arise, and a superposition of such operators may have a meaningful physical interpretation.

A superposition of boundary operators conditioned at the initial time reduces to an ordinary superposition of initial states. Post-selected decoherence functionals allow a richer structure, including possible entanglement between initial and final conditions. We leave the analysis of these possibilities to future work.

\subsection{Path-integral formulation}

Suppose that the single-time Hilbert space is ${\cal L}^2(Q)$, the space of square-integrable functions on a configuration space $Q$. Fine-grained configuration-space histories are then paths $q(\cdot)$ on $Q$, namely, maps from the interval $[0,T]$ to $Q$.

For such paths, the standard decoherence functional takes the form \cite{hartlelo}
\bey
d[q(\cdot),q'(\cdot)]=\psi_0[q(0)]\psi_0^*[q'(0)]\delta[q(T)-q'(T)]
e^{iS[q(\cdot)]-iS[q'(\cdot)]},
\eey
where $S[q(\cdot)]$ is the action functional.

A superposition of dynamics corresponds to a decoherence functional involving several action functionals $S_a[q(\cdot)]$,
\bey
d[q(\cdot),q'(\cdot)]=\psi_0[q(0)]\psi_0^*[q'(0)]\delta[q(T)-q'(T)]
\sum_{a,b}f_{ab}e^{iS_a[q(\cdot)]-iS_b[q'(\cdot)]},
\eey
with
\[
\sum_{a,b}f_{ab}=1.
\]
The off-diagonal coefficients $f_{ab}$, with $a\neq b$, encode interference between distinct action principles.

Suppose now that each action can be decomposed as
\[
S_a=S_0+F_a,
\]
where $S_0$ is common to all alternatives and $F_a$ contains the dependence on $a$. The decoherence functional may then be written in the form
\bey
d[q(\cdot),q'(\cdot)]=\psi_0[q(0)]\psi_0^*[q'(0)]\delta[q(T)-q'(T)]
e^{iS_0[q(\cdot)]-iS_0[q'(\cdot)]+W[q(\cdot),q'(\cdot)]}, \label{dfif}
\eey
where
\bey
W[q(\cdot),q'(\cdot)]=\log\sum_{a,b}f_{ab}
e^{iF_a[q(\cdot)]-iF_b[q'(\cdot)]}. \label{iiiff}
\eey
Note that the indices $a, b$ need not be discrete, they may be continuous, or even be elements of a space of paths.

The quantity $W[q(\cdot),q'(\cdot)]$ plays the role of an influence functional. Influence functionals were introduced by Feynman in the analysis of open quantum systems \cite{FeVe63}, and Eq. (\ref{dfif}) indeed has the structure of an open-system decoherence functional. Hence, superposition of dynamics may be equivalent to the dynamics of an open quantum system that is manifested even in the absence of a physical environment.

The expression (\ref{iiiff}), however, is more general than the standard influence functional obtained by tracing out an environment governed by ordinary quantum theory. Its only constraint is the positivity condition of $f_{ab}$ that guarantees positivity of the decoherence functional. Consequently, the histories axioms admit influence functionals that do not arise from the standard open-system construction.

\subsection{A histories quantum switch}

Some decoherence functionals involving superpositions of dynamics can be obtained from an ordinary decoherence functional on an enlarged composite system. The construction generalizes the mechanism underlying the quantum switch, in which a control subsystem correlates orthogonal control states with different ordered compositions of operations \cite{ChiribellaDArianoPerinottiValiron2013, Araujo15, Procopio15, Rubino17, Goswami18}.

Consider a composite system with history Hilbert space
\[
{\cal V}_1\uotimes{\cal V}_2,
\]
where
\[
{\cal V}_1=\bigotimes_i{\cal H}^{(1)}_{t_i}
\]
describes the system of interest and
\[
{\cal V}_2=\bigotimes_i{\cal H}^{(2)}_{t_i}
\]
describes the switch degrees of freedom. We use $\uotimes$ for tensor products between subsystems, reserving $\otimes$ for tensor products between distinct times. Assume that both the translation and boundary operators factorise,
\bey
{\cal S}={\cal S}^{(1)}\uotimes{\cal S}^{(2)},
\qquad
\hat{\cal A}_{im}=\hat{\cal A}_i^{(1)}\uotimes\hat{\cal A}_m^{(2)}.
\eey
Here $|i\rangle$ and $|m\rangle$ label orthonormal bases of the single-time Hilbert spaces ${\cal H}^{(1)}$ and ${\cal H}^{(2)}$, respectively.

Let the dynamics operator of the composite system be $\sum_a {\cal U}_a\uotimes  E_a$,
where the projectors $\hat E_a$ form an exclusive and exhaustive set on ${\cal V}_2$. The switch degree of freedom thereby correlates each projector $ E_a$ with a different system dynamics ${\cal U}_a$.

For histories of the form $\tilde\alpha\uotimes  I$ and $\tilde\beta\uotimes  I$, the reduced decoherence functional of the system becomes
\bey
d(\alpha,\beta)=\sum_i\sum_{a,b}f_{ab}
g_i^*\!\left( {\cal U}_a^\dagger\tilde\alpha {\cal U}_a\right)
g_i\!\left( {\cal U}_b^\dagger\tilde\beta {\cal U}_b\right),
\eey
where $f_{ab} = \sum_m h^*_m(a) h_m(b)$, with
\bey
h_m(a) = \operatorname{Tr}_{{\cal V}^{(2)}}\!\left( {\cal A}_m^{(2)}  E_a{\cal S}^{(2)}\right).
\eey
Thus, after the switch degrees of freedom are ignored, the system is described by a decoherence functional of the general form associated with a mixture of superpositions of dynamics.

For a factorised projector
\[
  E_a=  E_{a_{t_1}}\otimes\cdots\otimes  E_{a_{t_n}}
\]
and an initial switch state $|\phi_0\rangle$, one obtains
\bey
f_{ab}=\langle\phi_0|
  E_{b_{t_1}}  E_{b_{t_2}}\cdots  E_{b_{t_n}}E_{a_{t_n}}
\cdots  E_{a_{t_2}}  E_{a_{t_1}}
|\phi_0\rangle. \label{fab2}
\eey
At least a two-time history is required for $f_{ab}$ to possess non-diagonal components and hence to encode coherence between different dynamics.

This demonstrates that “superposition of dynamics” is not merely an exotic extension of quantum theory. Standard quantum mechanics can realize a subclass of it.
However, the matrix $f_{ab}$ of Eq. (\ref{fab2}) does not factorise as $f_{ab}=\lambda^*_a\lambda_b$.
A switch therefore generates, in general, a mixture of superpositions of dynamics rather than a pure superposition. 

A pure superposition arises  through post-selection on a final switch state $|\phi_f\rangle$ at some final time $T$. In that case,
\bey
f_{ab}=\frac{\langle\phi_0|
  E_{b_{t_1}}  E_{b_{t_2}}\cdots  E_{b_{t_n}}
|\phi_f\rangle
\langle\phi_f|
  E_{a_{t_n}}\cdots  E_{a_{t_2}}  E_{a_{t_1}}
|\phi_0\rangle}{|\langle \phi_f|\phi_0\rangle|^2}.
\eey
%Thus, quantum-switch-type controlled dynamics provides a particular realization of superposition of dynamics within an enlarged histories theory. More generally, tracing out the switch yields a mixture of superpositions, while a pure superposition requires the reduced coefficient matrix $f_{ab}$ 
% to have rank one, as occurs here upon suitable postselection.

\section{Measurements and operations}

In this section, we examine the implications of superpositions of dynamics in measurement set-ups---i.e., setups that involve  a physical system interacting with a measurement apparatus.
\subsection{A model for sequential measurements}

Consider first a system with Hilbert space ${\cal H}$ and a measuring apparatus with Hilbert space ${\cal K}$. For simplicity, we assume that the only contribution to time evolution is the system-apparatus interaction, given by $U = \sum_a E_a \uotimes V_a$. Here, the operators $E_a$ are projectors on ${\cal H}$, while the $V_a$ are unitary operators on ${\cal K}$. We assume that the apparatus is initially in the state $|\phi_0\rangle$ and that its possible measurement records are represented by an exclusive and exhaustive set of projectors $P_{\lambda}$ on ${\cal K}$. 

After tracing out the apparatus degrees of freedom, an initial system state $\rho_0$, conditioned on the outcome $\lambda$, is transformed into ${\cal M}_{\lambda}(\hat{\rho}_0)/\mbox{Prob}(\lambda)$, where
\bey
{\cal M}_{\lambda}(\rho_0) = \sum_{ab} u_{ab}(\lambda) E_a \hat{\rho}_0 E_b, \quad u_{ab}(\lambda) = \langle \phi_0|V_b^{\dagger}P_{\lambda}V_a|\phi_0\rangle, \label{mop}
\eey
while 
\bey
\mbox{Prob}(\lambda) =  Tr {\cal M}_{\lambda}(\rho_0) = \sum_a u_{aa}(\lambda) Tr( \rho_0 E_a ).
\eey

Consider next the same system interacting with two apparatuses, labelled by $r=1,2$. Apparatus $r$ is described by a Hilbert space ${\cal K}_r$, is prepared in an initial state $|\phi_0^{(r)}\rangle$, and interacts with the system through the unitary $U_r=\sum_{a_r}E^{(r)}_{a_r}\uotimes V^{(r)}_{a_r}$. Its possible records are represented by an exclusive and exhaustive set of projectors $P^{(r)}(\lambda^{(r)})$. 

If the system interacts first with apparatus 1 and then with apparatus 2, its unnormalised final state, conditioned on the records $\lambda_1$ and $\lambda_2$, is ${\cal M}^{(2)}[{\cal M}^{(1)}(\rho_0)]$,
and the associated probabilities are 

\bey
p_{1 \prec 2} (\lambda_1, \lambda_2) = Tr_{\cal H}  {\cal M}^{(2)}[{\cal M}^{(1)}(\hat{\rho}_0)]= \sum_{a_1 a_1'} \sum_{a_2a_2'} u^{(2)}_{a_2 a_2'}(\lambda_2) u^{(1)}_{a_1 a_1'}(\lambda_1)  Tr_{\cal H}(E^{(2)}_{a_2}E^{(1)}_{a_1} \rho_0 E^{(1)}_{a_1'} E^{(2)}_{a_2'} ). \label{p1<2}
\eey
Conversely, if the system interacts first with apparatus 2 and then with apparatus 1, its unnormalised final state is ${\cal M}^{(1)}[{\cal M}^{(2)}(\hat{\rho}_0)]$,
and the associated probabilities are
\bey
p_{2 \prec 1} (\lambda_1, \lambda_2) = \sum_{a_2 a_2'} \sum_{a_1 a_1'} u^{(1)}_{a_1 a_1'}(\lambda_1) u^{(2)}_{a_2 a_2'}(\lambda_2)  Tr_{\cal H}(E^{(1)}_{a_1} E^{(2)}_{a_2} \rho_0 E^{(2)}_{a_2'} E^{(1)}_{a_1'} ).
\eey

\subsection{Process matrices and indefinite causal order}

The preceding construction assumes that the two apparatuses interact with the system in a definite temporal order. The alternatives $(1\prec 2)$ and $(2 \prec 1)$ correspond to two different sequential compositions of quantum instruments,
\bey
{\cal M}^{(2)}(\lambda_2)\circ {\cal M}^{(1)}(\lambda_1),
\qquad
{\cal M}^{(1)}(\lambda_1)\circ {\cal M}^{(2)}(\lambda_2).
\eey
In the   circuit description of a measurement protocol, one must specify one of these orderings in advance. One may also allow a classical random choice between the two orderings. The resulting joint probabilities then take the form
\bey
p(\lambda_1,\lambda_2) = 
q p_{1\prec2}(\lambda_1,\lambda_2)
+
(1-q) p_{2\prec1}(\lambda_1,\lambda_2),
\eey
for some $0 \leq q \leq 1$. Such a description permits uncertainty about the order, but still presupposes that each individual realisation has a definite ordering.

The process-matrix formalism was introduced to describe multipartite quantum experiments without assuming a definite global causal order from the outset \cite{OreshkovCostaBrukner2012}. The starting point is that standard quantum theory holds locally in each laboratory. Thus, each laboratory receives an input subsystem, implements an instrument, and produces an output system. No global rule is imposed at the outset for how the output of one laboratory is connected to the input of another. Instead, the joint probabilities for local instrument elements are written as
\bey
p(\lambda_1,\lambda_2)
=
\mathrm{Tr}\left[
W
\left(
{\cal M}^{(1)}(\lambda_1)\otimes
{\cal M}^{(2)}(\lambda_2)
\right)
\right],\label{WW}
\eey
where ${\cal M}^{(r)}(\lambda^{(r)})$ is viewed as an element of $({\cal H} \otimes {\cal H}^*)^2$ and $W$ is an element of $({\cal H} \otimes {\cal H}^*)^4$.

Ordinary quantum circuits are contained in this framework. A circuit in which the output of laboratory (1) is transmitted to the input of laboratory (2) defines a process matrix $W_{1 \prec 2}$; similarly, the reverse circuit defines $W_{2\prec1}$. In the bipartite setting considered here, a process is called causally separable when it can be expressed as a convex combination $
qW_{1\prec2}+(1-q)W_{2\prec1}$.

The significance of the formalism is that its consistency conditions admit process matrices that cannot be decomposed in this way. Such processes are termed causally nonseparable. In the original proposal, this was interpreted as allowing the  order between local operations to be indefinite. The authors exhibited abstract processes whose correlations violate causal inequalities: constraints satisfied by all correlations generated by a definite causal order, or by a classical mixture of definite orders \cite{OreshkovCostaBrukner2012}.  

%A distinct and more concrete motivation is provided by the quantum switch \cite{ChiribellaDArianoPerinottiValiron2013}. In this protocol, a control system correlates one branch of the evolution with the ordering ($1 \prec 2$), and another branch with the ordering ($2 \prec 1$). The order label is therefore itself a quantum variable and can participate in interference. The associated process may become causally nonseparable.

%The phrase ``indefinite causal order'' is therefore used in the process-matrix literature in a specifically operational sense: the observable statistics cannot be represented as a convex mixture of processes with the alternative definite orders. It does not, by itself, determine the ontological status of the events underlying the process. In particular, the quantum switch is causally nonseparable, but it does not violate a causal inequality \cite{Araujo15}. Its status as an instance of indefinite causal order is established through causal witnesses, which detect the impossibility of a causally separable decomposition.

\subsection{Relation to the decoherence functional}

The process-matrix formalism can be related directly to the decoherent-histories framework. To see this, note first that 
the term $Tr_{\cal H}(E^{(2)}_{a_2}E^{(1)}_{a_1} \hat{\rho}_0 E^{(1)}_{a_1'} E^{(2)}_{a_2'} )$ in Eq. (\ref{p1<2}) defines a decoherence functional for the system, and by the ILS theorem,  
\bey
Tr_{\cal H}(E^{(2)}_{a_2}E^{(1)}_{a_1} \hat{\rho}_0 E^{(1)}_{a_1'} E^{(2)}_{a_2'}) = Tr_{{\cal V} \otimes {\cal V}} \left[ X_{1\po 2} (E^{(1)}_{a_1} \otimes E^{(2)}_{a_2}) \otimes (E^{(1)}_{a_1'} \otimes E^{(2)}_{a_2'}) \right]\nonumber 
\eey
in terms of the ILS operator $X_{1 \po 2}$. Similarly,
\bey
Tr_{\cal H}(E^{(1)}_{a_1} E^{(2)}_{a_2} \hat{\rho}_0 E^{(2)}_{a_2'} E^{(1)}_{a_1'} ) = Tr_{{\cal V} \otimes {\cal V}} \left[ X_{2\po 1} (E^{(1)}_{a_1} \otimes E^{(2)}_{a_2}) \otimes (E^{(1)}_{a_1'} \otimes E^{(2)}_{a_2'}) \right]\nonumber 
\eey
in terms of the ILS operator $X_{2\po 1}$. 

Let $B: {\cal V} \otimes {\cal V} \rightarrow \otimes_i ({\cal H}_{t_i}\otimes {\cal H}_{t_i})$ be a unitary operator defined by 
\bey
B\left(|u_{t_1}\rangle \ldots \otimes |u_{t_n}\rangle \otimes |v_{t_1}\rangle \otimes \ldots \otimes |v_{t_n}\rangle\right) = 
(|u_{t_1}\rangle \otimes |v_{t_1}\rangle) \otimes \ldots \otimes (|u_{t_n}\rangle \otimes |v_{t_n}\rangle) .
\eey
It then follows that $p_{1 \prec 2}(\lambda_1,\lambda_2)$ has the form (\ref{WW}), with the process matrix $W$ related to the ILS operator $X$ by
\bey
W = BXB^{\dagger}. \label{pmils}
\eey
The key point is that the probabilities
\bey
p(\lambda_1, \lambda_2) =  \sum_{a_1 a_1'} \sum_{a_2a_2'} u^{(2)}_{a_2 a_2'}(\lambda_2) u^{(1)}_{a_1 a_1'}(\lambda_1) d(\alpha_{a_1, a_2}, \alpha_{a_1', a_2'}) \label{iii}
\eey
are well defined for any strongly positive decoherence functional $d(\alpha,\beta)$, and not only for the canonical form appearing in Eq. (\ref{p1<2}). Conversely, by linearity, any process matrix defines, after the reshuffling (\ref{pmils}), an ILS operator and hence an effective strongly positive decoherence functional.

Hence, relation (\ref{pmils}) also applies to decoherence functionals that are not of the canonical form. In particular, it applies to decoherence functionals involving superpositions of the translation operator ${\cal S}$. By the same arguments leading to Eq. (\ref{superpd}), their most general form is
\bey
 d(\alpha_{a_1, a_2}, \alpha_{a_1', a_2'})  = \frac{1+n_3}{2(1+n_1)} Tr_{\cal H}(E^{(2)}_{a_2}E^{(1)}_{a_1} \hat{\rho}_0 E^{(1)}_{a_1'} E^{(2)}_{a_2'}) + \frac{1-n_3}{2(1+n_1)} Tr_{\cal H}(E^{(1)}_{a_1} E^{(2)}_{a_2} \hat{\rho}_0 E^{(2)}_{a_2'} E^{(1)}_{a_1'} ) \nonumber \\
 +  \frac{n_1 - i n_2}{2(1+n_1)} Tr_{\cal H}(E^{(2)}_{a_2} E^{(1)}_{a_1} \hat{\rho}_0 E^{(2)}_{a_2'} E^{(1)}_{a_1'} ) 
 +  \frac{n_1 + i n_2}{2(1+n_1)} Tr_{\cal H}(E^{(1)}_{a_1} E^{(2)}_{a_2} \hat{\rho}_0 E^{(1)}_{a_1'} E^{(2)}_{a_2'} ), \hspace{0.5cm}
\label{iii2}
\eey
for $|{\bf n}| \leq 1$. 

CP maps of the form (\ref{mop}), constructed from mutually commuting projectors $E_a$, are not the most general ones. General CP maps can be constructed by composing maps of the form (\ref{mop}). For example, suppose that, prior to the measurement, the system is coupled to an environment through a controlled interaction of the form $\sum_a F_a\otimes V_a$, where $F_a$ are projectors. The effective CP map for the first measurement is ${\cal M}^{(1)}(\lambda_1) \circ {\cal N}$, where $$ {\cal N}[\rho] = \sum_{ab} v_{ab} F_a \rho F_b,$$
for suitable coefficients $v_{ab}$. By varying the projectors $E_a$ and $F_a$, one can implement an arbitrary CP map for each measurement.
  
We therefore conclude that, modulo reshuffling, a process matrix is the ILS operator associated with the effective strongly positive decoherence functional of the system degrees of freedom. The demonstration here was for the case of two measurements. The structure of the construction gives strong reason to expect that the correspondence extends to an arbitrary number of measurements.

\subsection{Connection with superposition of dynamics}
We now consider the converse construction: whether a decoherent-histories analysis of sequential measurements leads naturally to process matrices. 
 
Because the pointer values of the two apparatuses can be specified simultaneously, the process could in principle be represented by a single-time history. However, it is convenient to work with a three-time history. Time $t_0$ corresponds to the initial condition, time $t_1$ to the first interaction, and time $t_2$ to the second interaction and measurement. 

The histories Hilbert space ${\cal V}$ is then
\bey
{\cal V} = \otimes_{i=0}^2 ({\cal H} \uotimes {\cal K}_1 \uotimes {\cal K}_2)_{t_i}.
\eey
The histories $\alpha_{\lambda}$ correspond to projectors $ I \otimes   I \otimes (I \uotimes  P^{(1)}_{\lambda_1} \uotimes P^{(2)}_{\lambda_2
})$. For brevity, we write $\lambda=(\lambda_1,\lambda_2)$.

The key observation is that the two possible temporal orderings define distinct dynamics operators:
  
\bey
 {\cal U}_{1 \po 2} = I\otimes U_1 \otimes U_2 U_1 \\
{\cal U}_{2 \po 1} = I\otimes U_2 \otimes U_1 U_2,
\eey
For the operators $U_1$ and $U_2$ introduced above, this gives 
\bey
d^{1 \prec 2}(\lambda, \mu) = \delta_{\lambda_1\mu_1} \delta_{\lambda_2\mu_2} p_{1 \prec 2} (\lambda_1, \lambda_2)\\
d^{2 \prec 1}(\lambda, \mu) = \delta_{\lambda_1\mu_1} \delta_{\lambda_2\mu_2} p_{2 \prec 1} (\lambda_1, \lambda_2)
\eey
Let us denote the orderings $1\po2$ and $2\po1$ by $\uparrow$ and $\downarrow$, respectively. A decoherence functional describing a superposition of these dynamics is
\bey
d(\lambda, \mu) = \sum_{ab} f_{ab} \langle \Psi| Q_b(\mu) Q_a(\lambda) |\Psi\rangle,
\eey
where $|\Psi\rangle$ is the initial state of the composite system, and
\bey
Q_{\uparrow}(\lambda) = U^{\dagger}_1 U^{\dagger}_2 (I \uotimes \hat{P}^{(1)}_{\lambda_1} \uotimes \hat{P}^{(2)}_{\lambda_2
})U_2 U_1\\
Q_{\downarrow}(\lambda) = U^{\dagger}_2 U^{\dagger}_1 (I \uotimes \hat{P}^{(1)}_{\lambda_1} \uotimes \hat{P}^{(2)}_{\lambda_2
})U_1 U_2.
\eey
A measurement with definite outcomes corresponds to a decoherence functional that is diagonal in the pointer alternatives. The probabilities are then given by its diagonal elements:
 \bey
 p(\lambda) = d(\lambda, \lambda) = \sum_{abcd}\sum_{a'b'c'd'} u^{(1)}_{aa'cc'}(\lambda_1) u^{(2)}_{bb'dd'}(\lambda_2) D_{abb'a', cdd'c'},
 \eey
 where 
 $$u^{(i)}_{aa'cc'}(\lambda_i) = \langle \phi_0^{(i)}| V_a^{(i)\dagger} P^{(i)}(\lambda_i) V_{a'}^{(i)}V_c^{(i)\dagger} P^{(i)}(\lambda_i) V^{(i)}_{c'}|\phi_0^{(i)}\rangle,
$$
and $D_{abb'a',cdd'c'}$ can be expressed in terms of a decoherence functional for the system degrees of freedom:
\bey
D_{abb'a',cdd'c'} = f_{\tiny \uparrow \uparrow} \langle \psi_0|E_a^{(1)}E_b^{(2)}E_{b'}^{(2)}E_{a'}^{(1)}E_{c}^{(1)}E_{d}^{(2)}E_{d'}^{(2)}E_{c'}^{(1)}|\psi_0\rangle
\nonumber \\
 + 
f_{\tiny \downarrow \downarrow} \langle \psi_0|E_b^{(2)}E_a^{(1)} E_{a'}^{(1)}E_{b'}^{(2)}E_{d}^{(2)}E_{c}^{(1)}E_{c'}^{(1)}E_{d'}^{(2)} |\psi_0\rangle \nonumber \\
+ 2 \mbox{Re} \left[f_{\tiny\downarrow  \uparrow } \langle \psi_0|E_a^{(1)}E_b^{(2)}E_{b'}^{(2)}E_{a'}^{(1)}E_{d}^{(2)}E_{c}^{(1)}E_{c'}^{(1)}E_{d'}^{(2)}|\psi_0\rangle \right] \label{jdjd}
\eey
In general, $p(\lambda)$ is not a functional of two distinguishable CP maps ${\cal M}^{(1)}(\lambda_1)$ and ${\cal M}^{(2)}(\lambda_2)$ acting on the initial state of the system. Thus, it cannot be brought in the form (\ref{WW}).

 A superposition of dynamics for the full system plus apparatus does not in general lead to separable interventions.  The process-matrix description emerges only after a locality/separability condition on interventions.

To identify the separability conditions, we compare Eq. (\ref{jdjd}) to Eqs. (\ref{iii}) and (\ref{iii2}). The first two terms in Eq. (\ref{jdjd}) correspond to convex combinations of  $p_{1 \prec 2} $ and $p_{2 \prec 1}$ and they are essentially the same. However, the third term of Eq. (\ref{jdjd}) 
involves the expectation value of a product of six operators  
$$\langle \psi_0|E_a^{(1)}E_b^{(2)} E_{a'}^{(1)}E_{d}^{(2)}E_{c}^{(1)} E_{d'}^{(2)}|\psi_0\rangle,$$ 
 while the third term of Eq. (\ref{iii2}) involves the product of four operators
 $$\langle \psi_0|E_a^{(1)}E_b^{(2)} E_{a'}^{(1)}  E_{b'}^{(2)}|\psi_0\rangle. $$
 
Hence, the existence of distinguishable operations, through which to identify a process matrix, requires a factorization condition. Writing $F_A = E_a^{(1)}E_b^{(2)}$ with $A$ the composite index $(a, b)$, one sufficient    condition is
\bey
\langle  F_A F_B F_C \rangle =   K_1 \delta_{AB}  \langle     F_A  F_C  \rangle +  K_2 \delta_{BC} \langle  F_A  F_C  \rangle + K_3 \delta_{AC} \langle  F_A  F_B  \rangle,
\eey
for   constants $K_1, K_2,$ and $K_3$. 

When the factorization condition is satisfied, the probabilities become a convex combination of terms of the form (\ref{iii}), leading to a relation of the form (\ref{pmils}) between the ILS operator and the process matrix.

 \section{Indefinite order of events}
 
 In this section, we analyze the natural notion of an event in the histories formalism. In particular, we adapt the results of Ref. \cite{AnPl23} in constructing probabilities for the causal order of events in the present context.  We identify an event with the emergence of a definite macroscopic record of observation. This definition reflects standard usage in physics, for example, in quantum optics \cite{QOp}, high-energy physics \cite{GriHEP}, or quantum measurement and control \cite{WiMi}. 
 
\subsection{Single events}

 By identifying a quantum event with the emergence of a macroscopic record of observation, we can develop a formalism of quantum events leading to a probability distribution for the time of the event's occurrence. This is the Quantum Temporal Probabilities (QTP) approach to measurements \cite{QTP1,  QTP4}.
  In this framework, we consider a composite physical system that consists of a microscopic and a macroscopic component. The microscopic component is the quantum system to be measured and the macroscopic component is the measuring device.

 Let ${\cal H}$ be the single-time   Hilbert space of the composite system.   A measurement event is defined 
 with respect to a split of ${\cal H}$ in two subspaces: in ${\cal H}_+$ a definite macroscopic record of detection exists; in ${\cal H}_-$ it does not.  We
denote  the projection operator onto ${\cal H}_+$ as $P$ and the projector onto ${\cal H}_-$ as $Q:= 1  - P$.  

 Let ${\cal V} =  \otimes_{i=1}^n {\cal H}_{t_i}$ be the Hilbert space for $n$-time histories. The   histories 
 \bey
 \alpha_i   &:&= 
Q^{\otimes(i-1)}\otimes P\otimes I^{\otimes(n-i)}, \;\; \mbox{where}\; i=1, \ldots n \nonumber \\
\alpha_{\emptyset} &:&= Q^{\otimes n}, 
 \eey
are mutually exclusive and they are exhaustive since $\sum_{i=1}^n \alpha_i + \alpha_{\emptyset} = I^{\otimes n}$. 

Each history $\alpha_i$ corresponds to the proposition that the detection record first emerged at time $t_i$, so it makes reference to a specific time of events. The history $\alpha_{\emptyset}$ corresponds to the proposition that  no record emerged. For the canonical decoherence functionals,
$$
d(\alpha_i, \alpha_j) = Tr(C_{\alpha_i}\rho_0C_{\alpha_j}^{\dagger}),
$$
where $C_{\alpha_i} = e^{iHt_1} (Qe^{-iH\delta t})^{i-1}Pe^{-iH (t_n-t_i)}$. For simplicity, we assumed that $\delta t = t_{i+1}-t_i$ is constant. For appropriate macroscopic apparatuses, these histories decohere, and a careful implementation of the continuous limit leads to a general formula for a probability density $P(t)$ for the time of occurrence of a single event.

\subsection{Two events}
Consider now the case of two events. This may correspond to a measured system interacting with two distinct apparatuses, as in the previous section, but such a specification is not necessary. It suffices that the single-time Hilbert space \({\cal H}\) of the total system splits as
$
{\cal H}
=
{\cal H}_{-}\oplus{\cal H}_{1}\oplus{\cal H}_{2}\oplus{\cal H}_{12}.
$
Here ${\cal H}_{-}$ corresponds to no macroscopic record, ${\cal H}_{1}$ to a record only in the first apparatus, ${\cal H}_{2}$ to a record only in the second apparatus, and ${\cal H}_{12}$ to records in both apparatuses. Let the corresponding mutually orthogonal projectors be $Q,P_1,P_2$, and $P_{12}$, respectively.

It is also useful to introduce the projectors
$$
R_1=P_1+P_{12},
\qquad
R_2=P_2+P_{12},
$$
which represent the propositions that the first and second records, respectively, have occurred. These commute and satisfy $R_1R_2=P_{12}$. \footnote{Let \({\cal H}={\cal H}_S\otimes{\cal K}_1\otimes{\cal K}_2\), where \({\cal H}_S\) is the Hilbert space of the measured system and \({\cal K}_i\) are the Hilbert spaces of the two apparatuses, \(i=1,2\). Let \({\cal K}_i={\cal K}_{i-}\oplus{\cal K}_{i+}\), where the first subspace corresponds to no measurement record and the second to a definite measurement record. Then
\[
{\cal H}_{-}={\cal H}_S\otimes{\cal K}_{1-}\otimes{\cal K}_{2-},
\]
\[
{\cal H}_{1}={\cal H}_S\otimes{\cal K}_{1+}\otimes{\cal K}_{2-},
\qquad
{\cal H}_{2}={\cal H}_S\otimes{\cal K}_{1-}\otimes{\cal K}_{2+},
\]
and
\[
{\cal H}_{12}={\cal H}_S\otimes{\cal K}_{1+}\otimes{\cal K}_{2+}.
\]
}

On the history Hilbert space ${\cal V}$, we define an exclusive and exhaustive set with the following histories:
\begin{itemize}
\item $\alpha_{ij} = \ Q^{\otimes i-1} \otimes P_1^{\otimes j-i}\otimes P_{12} \otimes I^{\otimes(n-j)}$, for $i < j$,
\item $ \alpha_{ij} = \ Q^{\otimes j-1} \otimes P_2^{\otimes i-j}\otimes P_{12} \otimes I^{\otimes(n-i)}$,  for $i > j$,
\item $\alpha_{ii} = Q^{\otimes i-1} \otimes    P_{12}\otimes I^{\otimes(n-i)}$,
\item $\alpha_{\emptyset i} = Q^{\otimes i-1} \otimes    P_{2}^{\otimes(n-i+1)}$,
\item $\alpha_{i \emptyset} = Q^{\otimes i-1} \otimes   P_{1}^{\otimes(n-i+1)}$,
\item $\alpha_{\emptyset \emptyset} =  Q^{\otimes n}$.
\end{itemize}
The histories $\alpha_{ij}$ correspond to the first detection event occurring at time $t_i$ and the second at time $t_j$; the histories $\alpha_{\emptyset i}$ correspond to the first record never occurring and the second occurring at time $t_i$; the histories $\alpha_{i \emptyset}$ correspond to the second record never occurring and the first occurring at time $t_i$; and $\alpha_{\emptyset \emptyset}$ correspond to no record ever occurring. 

Again we can construct decoherence functionals of the canonical form in terms of class operators
\bey
C_{\alpha_{ij}} = e^{iHt_1} (Qe^{-iH\delta t})^{i-1}(P_1e^{-iH\delta t})^{j-i}P_{12}e^{-iH (t_n-t_j)}, \;\; i\leq j \nonumber \\
C_{\alpha_{ij}} = e^{iHt_1} (Qe^{-iH\delta t})^{j-1}(P_2e^{-iH\delta t})^{i-j}P_{12}e^{-iH (t_n-t_i)},\;\; i \geq j \nonumber \\
C_{\alpha_{\emptyset i}}
=
e^{iHt_1}
\left(Qe^{-iH\delta t}\right)^{i-1}
\left(P_2e^{-iH\delta t}\right)^{n-i}
P_2 e^{-iHt_n}, \nonumber \\
C_{\alpha_{i\emptyset}}
=
e^{iHt_1}
\left(Qe^{-iH\delta t}\right)^{i-1}
\left(P_1e^{-iH\delta t}\right)^{n-i}
P_1 e^{-iHt_n}.
\eey
From these operators we can construct probability densities $P(t_1, t_2)$ for the time of occurrence of the two events.

A key point here is that we can define a set of mutually exclusive and exhaustive history propositions for the ordering of two events:
\begin{itemize}
\item  for event 1 being earlier than event 2: $\alpha_{1\po 2} = \sum_{i, j: i < j} \alpha_{ij} + \sum_i \alpha_{i\emptyset}$,
\item  for event 2 being earlier than event 1: $\alpha_{2\po 1} = \sum_{i, j: j < i} \alpha_{ij} + \sum_i \alpha_{\emptyset i}$,
\item for events 1 and 2 being simultaneous: $\alpha_{1\sim 2} = \sum_i \alpha_{ii} $,
\item for no events occurring: $\alpha_{\emptyset \emptyset}$.
\end{itemize}
These histories are defined with direct reference to the ordering relation intrinsic to the space of temporal supports $ Sup$. For physical systems evolving in a background spacetime, this ordering relation reflects the spacetime causal structure. These histories are defined purely at the kinematical level. Their counterparts for continuous-time histories are expected to be invariant under time-translations generated by the Liouville operator (\ref{Liouville}).

A generalization to histories corresponding to the different causal orderings for $n$ events is straightforward.

By linearity, the  class operators associated to the histories of event ordering are
\bey
C_{\alpha_{1\po 2}} = \sum_{i, j: i < j} C_{\alpha_{ij}} + \sum_i C_{\alpha_{i \emptyset}}, \nonumber \\
C_{\alpha_{2\po 1}} = \sum_{i, j: i > j} C_{\alpha_{ij}} + \sum_i C_{\alpha_{\emptyset i}}, \nonumber \\
C_{\alpha_{1\sim 2}} = \sum_i (Qe^{-iH\delta t})^{i-1}P_{12}e^{-iH (t_n-t_i)}.
\eey

For a decoherent set of histories, we can define  a POVM for the ordering of the two events, with positive operators $\Pi_{\alpha} = C_{\alpha}C_{\alpha}^{\dagger}$. 
Consider a pure state $|\psi\rangle = c_{1\po 2} |\psi_{1\po 2}\rangle + c_{2\po 1} |\psi_{2\po 1}\rangle  $, 
such that $\Pi_{1\po 2} |\psi_{1\po 2}\rangle \simeq |\psi_{1\po 2}\rangle$ and $\Pi_{2\po 1} |\psi_{2\po 1}\rangle \simeq |\psi_{2\po 1}\rangle$. The state $|\psi\rangle$   involves a superposition of different orderings of events. 

This is a  notion of indefinite temporal/causal ordering, which refers to the ordering of the occurrence of measurement events. It is defined at the level of history propositions, while the indefinite order of events suggested from the process matrix formalism refers to the dynamics entering the decoherence functional. This distinction reflects the more fundamental distinction between two types of time transformation, in the dual-time histories.

\section{Discussion and outlook}
We showed that histories theory admits coherent superpositions of dynamics at the level of the decoherence functional. In measurement settings, under appropriate factorization conditions, these superpositions give rise to non-trivial process matrices. In the set-ups considered in this paper, process-matrix ICO is an operationally restricted realization of the more general phenomenon of superposed dynamics. The structure of the construction gives strong reason to expect that this relation extends to the multipartite case. Conversely, generic superpositions of dynamics need not factorize into distinguishable interventions and therefore need not admit a process-matrix representation.

Histories theory also distinguishes a conceptually different notion of indefinite order, associated with the causal ordering of physical events. We reserve the term ``event'' for the emergence of a definite macroscopic record and use ``intervention'' for an operation performed on a quantum system. This separates ICO of events, where causal order is itself a history observable, from ICO of interventions, where indefiniteness concerns the ordering of dynamical interactions. The distinction reflects the separation betwen he  kinematical and dynamical roles of time that is intrinsic to the histories framework.

A fuller analysis of these distinctions requires an extension to relativistic spacetime. Relativistic quantum theories encode spacetime symmetries through representations of the Poincar\'e group, while measurements must correspond to localized physical processes rather than abstract circuit slots. Such an extension calls for continuous-time histories and their relativistic generalizations \cite{Sav02,Sav03}, together with genuinely relativistic frameworks for quantum measurement. The Quantum Temporal Probabilities framework \cite{QTP1,QTP4}, as well as other approaches to relativistic quantum measurement \cite{HeKr, OkOz, FeVe, GGM22}, provides a natural setting in which to investigate how process matrices, decoherence functionals, and superpositions of dynamics arise in relativistic quantum field theory.

The framework developed here also raises broader questions concerning the physical realization of superposed dynamics. In the examples considered, the alternatives correspond to different compositions or orderings of interactions. An important open question is whether superposed dynamics can arise intrinsically, rather than through engineered control or post-selection. Any setting in which no unique Hamiltonian can be identified upon quantization is a potential candidate. This may include some formulations of quantum gravity, as well as quantum field theory in nonstationary background spacetimes. The analysis of Sec.~3.3 further suggests that fundamental superpositions of dynamics could provide a mechanism for generating modified quantum dynamics at low energies \cite{BLSSU}.

Addressing these possibilities requires a more detailed theory of measurement for systems with superposed dynamics. Single-time alternatives need not decohere automatically, as they typically do under ordinary quantum dynamics, so one must identify the relevant coarse grainings and measurement models under which operational probabilities emerge. This would make it possible to ask whether superpositions of dynamics can produce observable signatures that distinguish them both from ordinary unitary evolution and from effective open-system dynamics arising from unitary evolution on an enlarged system.

 \section*{Acknowledgements}
 K.S. acknowledges support from the Andreas Mentzelopoulos Foundation through the  Branded Academic Position on ``Quantum Science and Technology" at the University of Patras.
 Both authors acknowledge the COST Action CA23115 ``Relativistic Quantum Information".

\begingroup
\small
\setlength{\itemsep}{1em}

\endgroup
\end{document}